# The use of primary energy factors and $CO_2$ intensities - reviewing the state of play in academic literature

Sam Hamels[a,*], Eline Himpe[b], Jelle Laverge[b], Marc Delghust[b], Kjartan Van den Brande[b], Arnold Janssens[b] and Johan Albrecht[a]

[a] Department of Economics, Faculty of Economics and Business Administration, Ghent University, Sint-Pietersplein 6, B-9000 Gent, Belgium

[b] Department of Architecture and Urban Planning, Faculty of Engineering and Architecture, Ghent University, Jozef Plateaustraat 22, B-9000 Gent, Belgium

**Abstract**

Reaching the 2030 targets for the EU's primary energy use (PE) and $CO_{2eq}$ emissions (CE) requires an accurate assessment of how different technologies perform on these two fronts. In this regard, the focus in academia is increasingly shifting from traditional technologies to electricity consuming alternatives. Calculating and comparing their performance with respect to traditional technologies requires conversion factors (CFs) like a primary energy factor and a $CO_{2eq}$ intensity. These reflect the PE and CE associated with each unit of electricity consumed. Previous work has shown that the calculation and use of CFs is a contentious and multifaceted issue. However, this has mostly remained a theoretical discussion. A stock-taking of how CFs are actually calculated and used in academic literature has so far been missing, impeding insight into what the contemporary trends and challenges are. Therefore, we structurally review 65 publications across six methodological aspects. We find that 72% of the publications consider only a single country, 86% apply a purely retrospective perspective, 54% apply a yearly temporal resolution, 65% apply a purely operational (instead of a life-cycle) perspective, 91% make use of average (rather than marginal) CFs, and 75% ignore electricity imports from surrounding countries. We conclude that there is a strong need in the literature for a publicly available, transparently calculated dataset of CFs, which avoids the shortcomings found in the literature. This would enable more accurate and transparent PE and CE calculations, and support the development of new building energy performance assessment methods and smart grid algorithms.

**Highlights:**

- Primary energy factors and $CO_2$ intensities are crucial but contentious parameters
- We review how these 'conversion factors' are currently being calculated and used
- Across 65 recent publications, we structurally assess six methodological aspects
- Methodological shortcomings and other overarching challenges are identified
- A publicly available dataset of conversion factors would benefit future research

* Corresponding author. Email address: Sam.Hamels@UGent.be

## Nomenclature

| | |
|---|---|
| BEP | Building energy performance |
| CE | $CO_{2eq}$ emissions expressed in (kilo)grams (e.g. associated with a particular appliance, building (stock) or EV (stock)) |
| CEN | The European Committee for Standardization |
| CF | Conversion factor (i.e. PEF or CI) |
| CHP | Combined heat and power (plant) |
| CI | $CO_{2eq}$ intensity of electricity, expressed in $g/kWh_E$ |
| CoM | Covenant of Mayors |
| DHN | District heating network |
| DSM | Demand side management |
| EEA | European Environmental Agency |
| EED | Energy efficiency Directive (of the European Union) |
| EIA | Energy Information Administration (of the United States) |
| ELCD | European Reference Life Cycle Database |
| ENTSO-E | European Network of Transmission System Operators for Electricity |
| EPBD | Energy performance of buildings Directive (of the European Union) |
| ETS | Emissions trading system |
| EU | European Union |
| GCB | Gas condensing boiler |
| GEMIS | Global Emissions Model for integrated Systems |
| GHG | Greenhouse gas |
| GS | Geographical scope |
| HP | Heat pump |
| IEA | International energy agency |
| IINAS | International Institute for Sustainability Analysis and Strategy |
| IPCC | Intergovernmental panel on climate change established by the UNFCCC |
| $kWh_E$ | Kilowatt hour of electricity |
| $kWh_F$ | Kilowatt hour of fuel as measured by the primary energy stored in its chemical bonds |
| $kWh_P$ | Kilowatt hour of primary energy |
| LCA | Life-cycle analysis |
| LCP | Life-cycle perspective |
| MILP | Mixed integer linear programming |
| MPC | Model predictive control |
| MRIO | Multi-regional input-output (model) |
| NZEB | Nearly-zero energy building |
| OP | Operational perspective |
| PE | Primary energy (use) expressed in $kWh_P$ (e.g. associated with a particular appliance, building (stock) or EV (stock)) |
| PEF | Primary energy factor for electricity, expressed in $kWh_P/kWh_E$ |
| $PEF_F$ | Primary energy factor for a specific fuel, expressed in $kWh_P/kWh_F$ (cf. footnote 2) |
| $PEF_T$ / $CI_T$ | Primary energy factor or $CO_{2eq}$ intensity for a specific electricity generation technology, expressed in $kWh_P/kWh_E$ |
| PPA | Power purchasing agreement |
| PV | (Solar) photovoltaics |
| Ref(s) | Reference(s) |
| SEAP | Sustainable Energy Action Plan |
| SGA | Smart grid algorithm |
| TES | Thermal energy storage |
| TR | Temporal resolution |
| TS | Temporal scope |
| TSO | Transmission system operator |
| UCED | Unit commitment economic dispatch (model) |
| UNFCCC | United Nations Framework Convention on Climate Change |
| VRES | Variable renewable energy sources (wind and solar energy) |

# 1. Introduction

The European Union strives for sharp reductions in both its primary energy use (PE) and $CO_2$ equivalent emissions (CE), as formalized in the flagship 2030 policy targets. Electricity use represents a major component of PE and CE. In fact, its importance is continually increasing due to the ongoing electrification of the heating and transport sectors[1]. This highlights the need for an accurate evaluation of electricity consuming technologies. For example, when we calculate the CE associated with the electricity consumption of a heat pump (HP) or electric vehicle (EV). This requires a $CO_2$ equivalent intensity (CI), expressed in $gCO_{2eq}/kWh_E$). Similarly, calculating the PE associated with a particular electricity consumption requires a primary energy factor (PEF)[2], expressed in $kWh_P/kWh_E$. CIs and PEFs are sometimes called conversion factors (CFs), respectively converting an amount of electrical energy into either an amount of $CO_{2eq}$ emissions or primary energy.

CFs play a crucial role in determining the merits of different technologies and policy measures. Electricity consuming technologies like HPs and EVs become more attractive relative to their fossil fueled alternatives, when CFs are 'low'. At the same time, 'high' CFs improve the attractiveness of electricity-saving measures. For a given cost, more CE and PE will be saved. In other words, CFs co-determine the 'abatement costs' associated with different technologies and measures. They thereby co-determine the contents of the cost-optimal mix of technologies and measures, which can be identified and strived towards by policy makers.

Consequently, CFs can have a significant market impact. Favorably evaluated technologies like HPs (i.e. in the case of low CFs) could receive a higher amount of government support, increasing their market share vis-à-vis competing technologies. Similarly, CFs can have a significant impact on the proliferation of solar PV. For this technology, a high PEF is beneficial since the yearly electricity production from a building's solar installation is typically multiplied by the PEF. The resulting PE is subsequently subtracted from the building's overall PE, making PV an attractive option to (help) reach the PE target imposed by building energy performance policies.

Given all of the above, CFs have received a large amount of attention in academic literature. Across a wide range of studies in which CFs are mentioned, the importance of these parameters is frequently confirmed. For example, Refs [1,2] simulate several renovation scenarios for a range of buildings to identify cost-optimal solutions. In doing so, they find that the assumed PEF has a significant impact on the choice of renovation measures and heating technologies. Other Refs [3–11] draw similar conclusions with respect to the impact of CI assumptions. Some authors have even found that differences in the assumed CIs lead to completely opposing results, for example with respect to the decarbonization merits of EVs [8,9].

[1] The heating sector can be widely conceived of as all forms of heating in buildings and industry. However, our focus in this respect lies on the electrification of heating in buildings, through the proliferation of heat pumps.

[2] We use the term 'primary energy factor' as a synonym for what is sometimes more specifically called the 'primary energy factor for electricity'. The latter expression is used when separate primary energy factors are also defined for energy carriers like biomass, coal or natural gas. In those cases, the primary energy contained in the chemical bonds of the fuel itself needs to be conceptually separated from the total primary energy use that can be associated with consuming a particular amount the fuel. The latter can be higher than the former if the primary energy associated with extracting, processing and transporting the fuel is taken into consideration. We therefore conceptualize 'fuel PEFs' (denominated as $PEF_F$) expressed in $kWh_P/kWh_F$. The numerator $kWh_P$ stands for the total amount of primary energy (including extraction, processing, transport and the chemical energy in the fuel itself), while the denominator purely represents the amount of fuel as measured by the primary energy stored in the chemical bonds of the fuel itself. Logically, a fuel PEF can only have a value higher than 1 when 'life cycle aspects' are taken into account.

Nevertheless, there is no clear consensus on how CF should specifically be calculated and used, even though there are a number of important methodological aspects to consider. Previous work focusing on the methodological aspects related to CFs has remained limited to theoretical discussions, foregoing a review of the methodological choices that are actually made in the literature [6,7,12–14]. We therefore formulate the research question: "*How are PEFs and CIs calculated and used in recent academic literature?",* in the context of the European energy and climate goals. By answering this question, we can identify the issues that are prevalent in the literature, propose our own solutions and synthesize those already proposed by others.

Answering our research question requires a 'structural stocktaking' of academic literature with respect to the calculation and use of CFs. For a total of 65 Refs, we structurally analyze several aspects of the CFs they use in their PE and CE calculations for various electrical energy demands (Table 1). These aspects are: the number of countries for which a calculation was performed (*geographical scope*), whether the calculations use yearly average values or rather a higher *temporal resolution*, whether the calculations only consider the present or also project into the future (*temporal scope*), whether or not electricity imports from other countries are taken into account (*import perspective*), whether the parameters were based on historical data or calculated in another way (*parameter source*), whether the calculation applied an operational or life-cycle perspective (*assessment boundary*) and whether average or marginal parameters were considered (*market perspective*). Each of these methodological aspects are explained further and discussed in detail in section 2, together with the various 'options' in the case of each aspect. For example, the methodological aspect "temporal resolution" can include options like "yearly" and "hourly".

Different attempts at differentiating these methodological aspects have resulted in slightly varying overviews. For example, the overviews found in previous academic work ([6,7,12–14], see Table 2) and in the standards developed by CEN (prEN17423) and ISO (52000 series)[15,16] differ slightly from each other. In our case, the selection of methodological aspects was determined in a bottom-up way, meaning it was derived directly from of our literature sample.

In addition to reviewing the main methodological aspects (section 2), we also identify several other features associated with CFs and how they are used in the scientific literature (section 3). For example, where the CFs used in the literature typically *come from*, the (lack of) *transparency* about CFs and the calculations they are used in, the connection with non-academic uses of CFs (i.e. in official assessments of building energy performance), and so forth. They further contextualize the use of CFs in scientific literature, and shed a light on associated challenges and opportunities. We conclude in section 4.

# 2. Structural literature review of PEF and CI methodological aspects

## 2.1. Literature selection process

Our literature selection process was organized as follows. First of all, Refs obviously needed to make use of a PEF, a CI, or both. Secondly, the sheer depth of the building energy and EV literatures prompted us to limit ourselves to *recent* publications. We focus on Refs that are published no earlier than 2018. This helps ensure that we are truly taking a snapshot of the current state-of-play in academic literature, ignoring the use of CFs in older work. Thirdly, we focus on publications that cover the European geographical area. Some exceptions to the second and third criteria were made, motivated by our overarching goal of selecting illustrative and relevant examples of the use of PEFs and CIs in academic literature.

A final selection criterium was that CFs were clearly used to calculate the PE and CE associated with *electrical* energy demands. Refs where only the PE and CE of non-electrical energy demands are estimated (e.g. Refs [17,18]), were excluded from our review. Refs considering both, were included. It should also be noted that the use of CFs is not always the core focus of each of the included Refs. In fact, in many cases CFs are only mentioned briefly. However, we do not focus on the other content and findings of the selected Refs, which are *out of scope* for the purposes of our literature review.

## 2.2. Overview of the literature sample

An overview of the 65 Refs – focusing on their use of CFs – is provided in Table 1, acting as a guide throughout sections 2 and 3. For the sake of clarity, the Refs in Table 1 are grouped into several categories. A few categories can be distinguished with respect to PE and CE associated with buildings. Another category groups Refs that calculate the CE associated with EVs. Finally, a residual category groups Refs that are purely focused on the *calculation of CFs* for particular geographical areas. These Refs typically do not 'apply' their calculated values in a PE or CE calculation. Similar to the aforementioned works that *theoretically discuss* the methodological aspects related to CFs (cf. Table 2), the Refs in the residual category of Table 1 do not perform a structural stocktaking of the recent academic literature. For each of the methodological aspects, Figure 1 provides an overview of how prevalent each of the options are in our literature sample.

*Table 1: Structural review of how PEFs and CIs are calculated and applied across academic literature*

| Ref | PD | CF | GS | TR | TS | IP | S | DA | AB | MP | What is calculated |
|---|---|---|---|---|---|---|---|---|---|---|---|
| **Single appliance level** | | | | | | | | | | | |
| [19] | '18 | CI | NL, UK, DE, FR | y | r, p | p | o | na | l | a | CE of a HP + PV + battery |
| [20] | '17 | PEF | IT | y | r, p | p | h, o | '14 | o | m | PE savings of replacing a GCB with a HP |
| [21] | '18 | PEF | IT | h[1] | r | p | h | '15-'16 | o | a | PE of a HP as compared to providing heating with a CHP through a DHN |
| [4] | '18 | PEF | IT | h[2] | r | p | h | '11-'16 | o | a | PE of a HP as compared to a GCB |
| [22] | ‘12 | CI PEF | 27[3] | y | r | p | h | ‘08 | o | a | PE and CE of heat recovery ventilation systems |
| [23] | ‘12 | CI | UK | h[4] | r | p | h | ’09-‘11 | l | a | CE of a PV-connected battery |
| [24] | ‘17 | PEF | DE | h | r | p | o | na | o | a | PE calculation of individual HP’s combined with TES |
| [11] | ‘18 | CI PEF | EU[5] | y | r, p | p | h, o | ’12,‘14 | o | a | CE and PE calculation of a HP compared to a GCB, in the context of seven European countries. |
| [25] | ‘19 | PEF | LU | y | r | p | o[6] | na | o | a | PE calculation for a ventilation system |
| [26] | ‘20 | CI | 10[7] | h | r | p | h | ‘18 | l | a | CE calculation for HPs in 10 European countries |
| [27] | ‘19 | CI | NO | h | r | c | h | ‘15 | l | a | CE calculation for a HP |
| **Individual building level** | | | | | | | | | | | |
| [28] | ‘18 | CI | US (8)[8] | h | r | p | h | ‘17 | o | m | CE of a residential building with a home battery |
| [29] | ‘18 | CI PEF | SE | y | r | p | o | na | o | m | PE and CE of a multi-family building In Sweden |
| [1] | ‘17 | PEF | FI | y | r | p | h | ‘15 | o | a | PE and CE of a single apartment building in Finland |
| [30] | ‘19 | PEF | IT | y | r | p | h | ‘15 | o | a | PE calculation for a multi-family building in Italy |
| [31] | ‘19 | PEF | IT | y | r | p | h | ‘14 | o | a | PE calculation for a single-family building in Italy |
| [10] | ‘19 | CI PEF | CZ | y | r | p | h | ‘08 | o | a | PE and CE calculation for a ‘reference building’ in Czechia |
| [32] | ‘19 | CI PEF | SE | y | r | p | h | ‘16 | l | a | PE and CE calculation for a single-family building in Sweden |
| [33] | ‘20 | CI PEF | SE | y | r | p | o | na | l | m | PE and CE calculation for a multi-family building in Sweden |
| [34] | ‘19 | PEF | IT | y | r | p | h | ‘14 | o | a | PE calculation for an office building in Italy |
| [35] | ‘18 | CI PEF | CH | h | r | c | h | ‘15 | l | a | PE and CE calculation for a building in Switzerland |
| [36] | ‘18 | CI | DK | h | r* | c | h | ’13-‘16 | o | a | CE calculation for a single-family building in Denmark |
| [37] | ‘17 | CI | DK | h | r | c | h | ‘15 | o | a | CE calculation for a single-family building in Denmark |
| [38] | ‘16 | CI | DK | h | r | c | h | ’13-‘14 | o | a | CE calculation for an apartment building in Denmark |
| [39] | ‘18 | CI | ES | h | r | p | h | ‘16 | o | a, m[9] | CE calculation for a multi-family apartment building in Spain |
| [40] | ‘16 | CI | FR | h | r | p | h | ‘13 | l | a | CE calculation for a single-family building in France |
| [41] | ‘19 | CI PEF | IT | y | r | p | h | u | o | a | CE calculation for a hotel building in Italy, connected to EVs |
| [42] | ‘19 | CI PEF | CH | h | r | c | h | ’16-‘18 | l | a | PE and CE calculation for a single-family building in Switzerland |
| **Multiple buildings level** | | | | | | | | | | | |
| [43] | ‘15 | CI PEF | RS | y | r, p | p | h | ‘13 | o | a | CE and PE of 20 Serbian buildings |
| [44] | ‘10 | CI PEF | na | y | r | p | o[10] | na | l | t[10] | CE and PE of five Swedish buildings Some buildings use electrical resistance heating, others use heat pumps. |
| [45] | ‘19 | CI | US (2)[11] | h | r | c | h | ’10-‘14 | o | m | CE of multiple non-residential buildings in the US before and after the installation of batteries |
| [46] | ‘16 | PEF | PT | y | r | p | o | na | o | a | PE of a range of Portuguese buildings |
| [2] | ‘19 | CI | SE | y | r | p | h | ’13-‘15 | o | a | CE of a range of Swedish buildings |
| [47] | ‘17 | CI<br>PEF | BE | h | r, p | p | h | ’14-‘15 | o | a | CE and PE of two Belgian buildings (one terraced, one detached) |
| **Municipality level** | | | | | | | | | | | |
| [48] | ‘13 | CI | ES | y | r | p | h | ‘05 | o | a | CE calculation for the city of Barcelona |
| [49] | ‘19 | CI | HR | y | r | p | h | ‘16 | o | a | CE calculation for the city of Zagreb |
| [50] | ‘18 | CI | IT | y | r | p | h | ‘11 | o | a | CE calculation for 16 municipalities in Italy |
| [51] | ‘19 | CI | IT | y | r | p | h | ‘16 | o | a | CE calculation for the municipality of Évora |
| [52] | ‘18 | CI | IT | y | r | p | h | ’06,’08 | o, l | a | CE calculation for the municipality of Licata |
| [53] | ‘16 | CI | ES | y | r | p | h | ‘06 | o | a | CE calculation for the city of Gerona |
| [54] | ‘16 | CI | HR | y | r | p | h | ‘12 | o | a | CE calculation for the municipality of Korkula |
| [55] | ‘16 | CI | GR | y | r | p | h | ‘07 | o | a | CE calculation for the municipality of Ptolemaida |
| [56] | ‘17 | CI | FI | h | r | p | h[12] | ‘11 | o | a | CE calculation for three residential buildings and two commercial buildings in Finland |
| **Building stock level** | | | | | | | | | | | |
| [57] | ‘19 | PEF | DK | y | p | p | o | na | o | a | PE of the 2050 Danish building stock |
| [58] | ‘19 | CI | 28 | y | r | p | h | ‘12 | o | a | CE of various national building stocks in 2012 and 2030 |
| [59] | ‘18 | CI | DE, ES, FR, SE, UK | y | r, p | p | h | ’09-‘12 | o | a | CE of a representative building stocks for five European countries |
| [60] | ‘18 | CI<br>PEF | CH | y | r | p | h | ‘15 | l | a | CE and PE of a synthetic building stock representing Switzerland |
| **Electric vehicles** | | | | | | | | | | | |
| [61] | ‘19 | CI | DE | h[13] | r, p | p | h, o | ‘16 | l | a | CE of electric busses in Germany |
| [62] | ‘15 | CI | BE | h | r | p | h | ‘11 | l | a | CE of EV’s in Belgium |
| [9] | ‘18 | CI | EU, 28 | y | r | c | h | ‘13 | l | a | CE of EV’s in 28 European countries |
| [63] | ‘15 | CI | IT | h[14] | r | c | h | ‘13 | o | a | CE of EV’s in Italy |
| [64] | ‘17 | CI | DE, FR | h | r | p | h | ‘13 | o | a | CE of EV’s in France and Germany |
| [65] | ‘18 | CI | CZ, PL, o | y | r, p | p | h | ‘15 | l | a | CE of EV’s in Poland and Czechia |
| [66] | ‘18 | CI | SI | y | r | p | h | ’09-‘14 | o | a | CE of EV’s in Slovenia |
| **Other Refs** | | | | | | | | | | | |
| [67] | ‘18 | CI<br>PEF | IT | h | r | p | h | ’12-‘17 | o | a | Calculation and analysis of hourly PEFs and CIs, based on historical data from Italian TSO |
| [68] | ‘17 | PEF | SE, EU | y | r | p | h | ‘14 | l | a | Novel calculation of the $PEF_T$ of nuclear, and its impact on the Swedish and European PEF |
| [69] | ‘19 | CI<br>PEF | IT | h | r | p | h | ’16,’17 | o | a | Calculation of hourly PEF and CI |
| [70] | ‘14 | PEF | 19[15] | y | r | p | h | ’07-‘09 | o | a | Methodological proposal to calculate national PEF values |
| [71] | ‘18 | CI PEF | CH | h | r | c | h | ‘15 | l | a | Calculation analysis of hourly PEFs and CIs for Switzerland |
| [72] | ‘18 | CI | DK, NO, SE | h | r | c | h | ‘16 | l | a | Calculation of hourly CIs for Denmark, Norway and Sweden |
| [73] | ‘19 | CI | 6[16] | h | r | c | h | ‘15 | l | a | Calculation of hourly CIs for six Northern-European countries |
| [74] | ‘18 | CI | FR | h | r | c | h | ’12-‘14 | l | a | Calculation of hourly CIs for France |
| [75] | ‘14 | CI | NO, EU | h[17] | r | c | o[17] | na | o | a, m | Calculation of yearly CIs for Europe and hourly CIs for Norway |
| [76] | ‘19 | CI | 28 | h | r | c | h | ‘17 | l | a | Calculation and analysis of hourly CIs for 28 European countries |
| [77] | ‘15 | CI | ES | h | r | c | h | ‘12 | o, l | a | Calculation and analysis of hourly CIs for Spain |

Note: **Ref** = reference, **PD** = publication date **CF** = conversion factor used, **GS** = geographical scope of the CF (code of geographical areas like countries or the entire European area, the number of countries (further referenced below), or another approach), **TR** = temporal resolution (yearly or hourly), **TS** = temporal scope (is the CF calculated retrospectively or prospectively), **IP** = import perspective (production or consumption perspective on the CFs, cf. section 2.8), **DA** = data age (historical year on which the CF is based), **S** = CF source (is the CF purely based on historical data or on another approach), **AB** = assessment boundary (is the CF calculated from an operational or a life-cycle perspective), **MP** = market perspective (is the CF calculated as a market average, a marginal value in the market, or is a specific technology assumed). Notation “**na**” in any aspect column indicates ‘not applicable’, while notation “**u**” indicates ‘unclear’. The “*****” symbol is used to refer (part of) the “What is calculated” note of a Ref to one or several of its aspects (e.g. MP). Table contents are limited to the PE and CE calculations associated with electrical loads in each Ref, ignoring other content and contributions of the respective Refs. For more information on the reasoning behind our categorization of Refs, see Appendix A. For supplementary notes on some of the Refs included in Table 1, see Appendix B.

[1]: PEF calculation is only partially based on hourly historical data. Yearly values are used for all thermoelectric technologies.
[2]: Same ‘partially hourly’ approach as [3] with respect to PEF calculation.
[3]: PEFs and CIs calculated for 27 European countries
[4]: CI calculated using a highly stylized representation of UK electricity system.
[5]: Building characteristics are adapted to the typical circumstances in each of the seven countries (DE, FI, GR, IT, NL, SE, and UK) but European average CI and PEF values used for each case.
[6]: PEF simply assumed by authors to be 2.7, without providing a source.
[7]: The included countries are AT, CH, DE, DK, FR, IE, IT, NL, PL and UK.
[8]: CI estimated on for eight US regional grids.
[9]: CIs are calculated from a marginal perspective, while PEFs are calculated from an average perspective.
[10]: PE and CE related to building electricity demand is calculated for several scenarios. In each scenario, electricity is assumed to be generated by one technology (e.g. coal). $PEF_T$ and $CI_T$ values for each technology are estimated by the authors.
[11]: Buildings are assumed to be located in two US markets.
[12]: Hourly CIs are partially based on historical electricity production data for renewables and nuclear. The electricity generation by technologies for which data was not available, is estimated with a capacity and dispatch optimization model to cover the residual demand.
[13]: CI in 2016 is calculated on the basis of quarter-hourly historical data from a German TSO. CEs in future years (2017-2028) are calculated on the basis of yearly average CIs out of a German governmental report which made a rudimentary projection of CIs.
[14]: CI is calculated on the basis of yearly historical data from the Italian TSO (Terna). Hourly CIs derived by assuming hourly ‘load levels’ (production) of the different electricity generation technologies, because of a lack of hourly historical production data for Italy at the time of writing.
[15]: PEFs calculated for nineteen OECD member countries: AT, CA, CH, DE, DK, ES, FI, FR, GR, IE, IT, JP, NL, NO, PL, PT, SE, and US, on the basis of historical data.
[16]: The countries for which CIs are calculated are DE, DK, FI, NL, NO and SE.
[17]: For 2010, the UCED model is also used (i.e. its CIs are not based on historical production data). Hourly marginal CIs are calculated only for Norway.

*Table 2: Other references discussing CF methodological aspects*

| Ref | PD | CF | Goal of Ref | Note |
|---|---|---|---|---|
| [12] | '16 | PEF | Fraunhofer study procured by the European Commission to review PEF, $PEF_T$ and $PEF_F$ calculation methods | Detailed discussion of all possible PEF calculation methodologies with respect to $PEF_T$ calculations for various technologies, GS, TR, TS, IP, S, AB and MP. *Not* a review of academic literature. |
| [6] | '19 | PEF | Overview of PEF, $PEF_T$ and $PEF_F$ calculation methods of the IPCC, IEA, EIA and European legislative framework (EED and EPBD) | Stresses the fact that there are technical, political and economic dimensions to PEFs, which explains the current absence of a consensus on calculation methods. *Not* a review of academic literature. |
| [13] | '17 | PEF | Evaluation of the use of PEFs at various TRs | Also investigates various methodological options for the application of PEFs on a building's own production – and injection into the grid – of solar power. |
| [7] | '18 | CI PEF | Overview of PEF and CI calculation methods | Specifically focusses on the aspects of temporal resolution and the choice between an average or marginal market perspective. |
| [14] | '18 | PEF | Calculation of $PEF_T$ values for various technologies | Uncertainties about the input parameters in each $PEF_T$ calculation (e.g. power plant thermal efficiencies etc.) are used to generate a *spread* of possible $PEF_T$ values for each technology. |

Note: for abbreviations, see Table 1.

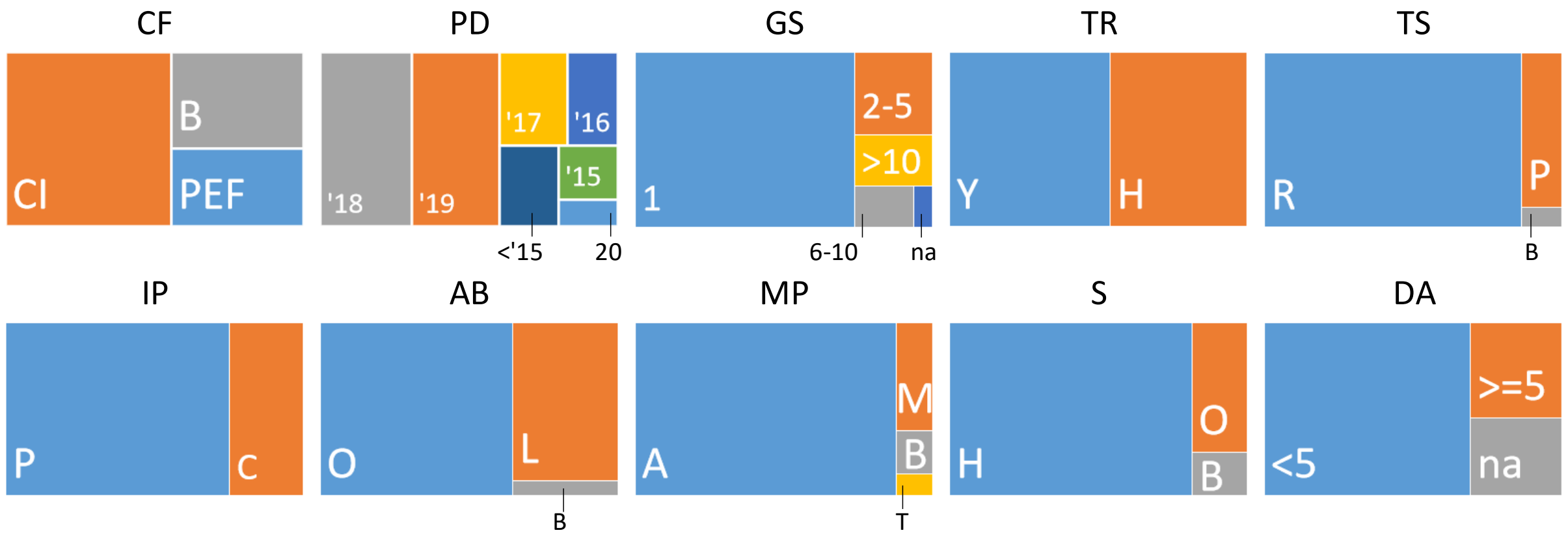

*Figure 1: Distributions found in our literature sample*

Note: for abbreviations, see Table 1. **GS** is shown here as the number of geographical areas considered. **DA** is shown here as the number of years between the Ref's publication date and the historical year to which the data used in the Ref refers. **B** stands for both.

## 2.3. Geographical scope

In our literature sample, 47 out of 65 (72%) Refs use CFs representing only a single country. Most of them ignore the fact that CFs are geographically variable and that their PE and CE results may therefore be highly specific to the single country they consider. In contrast, many Refs explicitly mention the importance of geographical variability [4,9,20,22,26,70,73,78]. Some of them do so to acknowledge the fact that their use of a single country's CFs is an important limitation. For example, Ref [4] calculates the PE associated with a HP and concludes that results are highly sensitive to the assumed PEF. The authors then call for *future* research to replicate their analysis within the context of other countries. Other Refs actually use geographically varying CFs themselves, and mention their importance to stress the added value of their work. For example, Ref [22] calculates the PE and CE for ventilation systems using CFs for 27 European countries. Similarly, Ref [9] estimates the CE associated with EVs using CIs for 28 European countries. Both Refs find that their results vary significantly across countries.

Crucially, these kind of approaches enable a *coherent comparison* of results across a variety of geographies. Such a comparison is not necessarily possible across separate studies. For example, across two studies that calculate the CE related to EVs in different countries. The methodologies and reference years used to calculate the CFs in each study may very well be different, resulting in an incoherent comparison. The various other ways in which CFs across different studies may differ are discussed in the following sections.

The fact that – so far – most studies have failed to use geographically diverse CFs, can be justified to a certain degree. Currently, it can require a significant amount of additional effort. First of all, because applying geographically diverse CFs may entail more than only changing the CF values themselves. Other modelling changes may be required as well. For example, if a building's PE or CE is calculated, both the characteristics of the building itself and the weather assumed in the simulations are likely to be representative for a particular country. To avoid inconsistencies, using geographically diverse CFs may then require changes to be made to those parts of the model as well. If such additional efforts are deemed excessive, authors may argue that considering the geographical variability of CFs lies 'out of scope' for their exercise.

However, the need to substantially change models can be avoided in some of these cases. For example, when other countries exist which have similar characteristics with respect to the other modelling aspects (i.e. similar buildings and weather in the aforementioned example), but different CFs. If this is the case, 'excessive additional effort' is not a compelling argument against using a variety of geographically diverse CFs and comparing the results.

Another reason why the use of geographically diverse CFs is associated with additional effort is the fact that it requires authors to gather or calculate multiple CFs. However, if up-to-date and transparently calculated CFs for various geographies would be made widely available to researchers, this problem could be severely limited (cf. section 3.1).

The single geographical area considered in most Refs is sometimes not a country but rather a region like Europe as a whole [11]. This raises a question about the difference in outcomes when using a European versus an individual country's CFs. However, to the best of our knowledge, none of the Refs in our literature sample perform the calculations required to make this comparison.

## 2.4. Temporal scope

56 out of our 65 Refs (86%) perform PE and CE calculations that are purely *retrospective* in nature. Their results could be significantly affected by the fact that CFs will continually change in the foreseeable future, as mentioned by a number of authors [4,6,20,24,38,40,47,70,75]. The main reason why *prospective* CFs are expected to differ, is because of the increasing penetration of renewables [4,20,24]. Other changes to the electricity mix are relevant as well, for example in the wake of the various phase-out decisions related to nuclear or fossil technologies. Prospective CFs are typically calculated by somehow simulating (scenarios for) the future electricity system.

The use of prospective CFs is valuable for two reasons, which we can explain using the calculation of the PE of a HP as an example. First of all, it allows for a comparison of the PE of a HP that is installed today versus one that is installed at some point in the future. Secondly, it allows the PE calculation for a HP that is installed today to take into account the changing PEF *throughout its operational lifetime*. Refs [6,24] show concretely that neglecting the expected decrease in the PEF can lead to a significant overestimation of the HP's PE. We can reasonably expect that the added value of prospective CFs could similarly be found in the case of other types of PE and CE assessments. For example, considering newly built projects, renovations, municipalities, entire national building stocks and EVs. However, for these applications, results comparing the use of retro- and prospective CFs are not yet available in the literature to the best of our knowledge.

A number of Refs explicitly consider future years in their analysis and yet assume (often implicitly) that the CFs they use – which are based on historical data – will remain constant over time. For example, Ref [30] optimizes the design of a multi-family building in Italy within a framework focusing on monetary costs and PE. The authors fail to mention the expected reductions of PEFs over time, even though they *do* include a sensitivity analysis testing the robustness of their 'optimal design' for future weather patterns as a result of climate change (which cause an increase in cooling needs).

Ref [58] calculates the CEs associated with various national building stocks for both the historical year 2012 and the future year 2030. The same national CIs for the year 2012 are assumed in the 2030 scenario. In other words, the analysis does not only ignore the sharp decline in CIs from 2012 until the Ref's publication date (2019), but even more importantly, it ignores the further declines expected in the years up to 2030.

Ref [59] calculates the CE of national building stocks of five European countries, before and after a range of different renovation packages. To calculate the CE associated with the building stocks' electricity consumption, they collect CIs based on historical data and *assume those to remain constant* over time (i.e. during the period in which the building-stock level renovation packages would be implemented, namely 2020-2050).

Of course, the challenge of generating prospective CFs is inherently complex and uncertain. Authors have dealt with this challenge in a variety of ways. In Ref [19], the analysis starts from retrospective CIs for four European countries on the basis of 2016 historical data, and subsequently generates very crude prospective CIs by linearly extrapolating them from their 2016 value down to zero in the year 2050.

Ref [20] calculates a retrospective PEF for Italy based on historical data for the year 2014, and generates a prospective PEF by creating a simplified scenario for the future Italian electricity system. Future wind and PV capacities are taken from the future scenarios made by ENTSO-E, and the future capacity factors of all other electricity generation technologies are roughly estimated on the basis of those renewable capacities.

Similarly, Ref [47] calculates prospective CFs through a stylized extrapolation of electricity production from variable renewable energy sources (VRES) to an assumed future value representing 40% of (historical) electricity demand. Meanwhile, the remaining part of demand is covered by a generic non-renewable technology, for which they assume a thermal efficiency of either 50% or 60%. This results in two prospective CFs for their future scenario.

Finally, Ref [75] uses an UCED model to simulate future scenarios for the European electricity system up to the year 2050, and calculates the European PEFs that result from those scenarios. The scenarios in question were made in the EU-funded project *SUSPLAN*, which was concluded in the year 2011 [79]. Although UCED simulations can be considered far superior to some of the aforementioned approaches to estimate future CFs, the scenarios in question are highly outdated in comparison with the state-of-the-art scenarios recently developed by ENTSO-E [80].

## 2.5. Temporal resolution

35 out of our 65 Refs (54%) exclusively make use of CFs with a yearly temporal resolution, ignoring temporal variations at the seasonal, monthly and even hourly level. However, a large number of Refs, including some of these 35, do mention the importance of a higher (in most cases hourly) temporal resolution [6,7,9,12,13,19,21,24,26–28,37,40,42,47,56,62,63,67,69,71–74,76,77,81,82]. A total of 30 Refs in our literature sample actually *apply* hourly CFs (cf. Table 1).

PE and CE calculations using a higher temporal resolution generate a *more accurate* result. In fact, many Refs have purposefully sought and found differences between what we may call the 'yearly' and 'hourly method' [4,21,24,26,40,42,56,69,77,81]. A PE and CE calculation is then performed once assuming the yearly values for the CFs and the respective electric energy demand, and another time using the hourly values.

The following are a few illustrative examples of such an analysis. Ref [26] analyses the difference between the outcomes of the yearly and hourly methods for HPs in ten European countries. The authors found that – in most cases – the difference lies in the order of magnitude of 5%. But for some combinations of HP consumption profiles and hourly CIs of particular countries, the difference can be as high as 10%. Similarly, Ref [56] compares the CE outcomes for two residential buildings and finds a difference of up to 6%. Finally, Ref [40] finds a rather remarkable difference of 30% between the outcomes of hourly and yearly methods, in their analysis of a highly energy efficient single family building in Switzerland. Each of these deviations effectively represent the under- and over-estimations that can result from applying yearly CFs. The literature shows that, whether a yearly CF under- or over-estimates the PE and CE is case-dependent.

It is important to reemphasize that the outcomes of PE and CE calculations can drive comparisons between different technologies. Deviations found between yearly and hourly methods are therefore more than an improvement in accuracy *for the sake of accuracy*. Applying a different temporal resolution can affect performance assessments of different technologies relative to each other. For example, Ref [4] compared HPs and GCBs and found that the former result in significant PE savings compared to the latter, but that the yearly method underestimated these savings by 8% when compared to the hourly method.

Hourly CFs are not generally available for use in academic research. The exceptions to this are the publications of real-time hourly CIs by the Danish TSO Energinet and the French TSO RTE. In the case of Energinet, three studies have already made use of this data [36–38]. To our best knowledge, no studies have so far been published that make use of the hourly CIs published by RTE. It should however be noted that the French electricity system – with its very high share of nuclear energy – is characterized by extremely low and rather stable hourly CIs. Therefore, the added benefit of applying an hourly method may be somewhat limited in the context of France.

Given the fact that hourly CIs for countries other than Denmark and France are unavailable, most Refs that make use of hourly CIs have typically calculated them themselves on the basis of historical electricity production data for the country in question [26,27,35,39,40,42,47,62,64,67,69,71–74,77].

Hourly CFs are expected to become more volatile in the future, due to the increasing penetration of VRES [4,7,19,27,69]. This will likely increase the importance of hourly methods over time. Moreover, a significant correlation between the hourly electricity generation by VRES, hourly PEFs and hourly CIs has also been found in the literature [4,67]. In section 3.6, we further explain the importance of volatile hourly CFs in the context of the development of smart grid algorithms (SGA's).

## 2.6. Assessment boundary

42 out of our 65 Refs (65%) consider CFs from a purely *operational perspective* (OP). Meaning, the CFs are only meant to represent the PE and CE associated with the act of generating electricity itself. The $PEF_T$ and $CI_T$ values for each electricity generation technology are then based on their operation alone. This enables certain technologies to have a $PEF_T$ of 0 $kWh_P/kWh_E$ (wind, solar and hydro) or a $CI_T$ of 0 $g/kWh_E$ (all renewables and nuclear). When an OP is used, the only thing that is potentially considered in addition to the generation of electricity itself, is its transportation from producer to consumer and the associated grid losses.

The alternative for an OP is a *life-cycle perspective* (LCP). In a LCP, the PE and CE associated with the entire life-cycle of each electricity generator is included in the calculation. For example, the PE and CE associated with building and eventually deconstructing a powerplant can also be included. A LCP implies that $PEF_T$ and $CI_T$ values can no longer be zero for certain technologies, because some amount of PE and CE will always be associated with non-operational aspects.

Two illustrative use cases for LCP CFs are life cycle analyses (LCAs) for buildings or EVs. Such analyses do not only consider the operational electricity consumption – for which LCP rather than OP CFs would then be used – but also the PE and CE embodied in materials use and other non-operational aspects. In a sense, LCP CFs are then used for the sake of 'consistently applying a life cycle perspective' across the entire analysis.

Assumptions about CFs can be highly important in LCAs, because they co-determine the ratio between operational and embodied PE and CE. This ratio is a crucial output variable of LCAs. After all, a relatively high share of embodied PE and CE is fundamentally what 'proves' the added value of any LCA when compared to using a simpler OP.

## 2.7. Market perspective

59 out of our 65 Refs (91%) make use of average CFs, which consider the entire mix of electricity generation from different technologies during a given timestep (e.g. a particular year or hour). An average CF is calculated as a weighted average, multiplying the electricity production by each technology (coal, wind, nuclear, etc.) with its respective $PEF_T$ or $CI_T$ value and dividing by the total aggregated electricity production.

An alternative for average CFs is to use marginal CFs, which only consider the marginal producer of electricity during a given timestep. This is the powerplant that would change its electricity production in response to an incremental change in electricity demand. A marginal CF does not need to be 'calculated', in the sense that it is simply equal to the $PEF_T$ or $CI_T$ of the marginal producer itself. Several

Refs claim that the choice between using average or marginal CFs can have a considerable impact on the outcome of PE and CE calculations [7,8,64].

To motivate the use of either kind of CF, a variety of arguments is used in the literature. Some Refs motivate their use of average CFs by arguing that the EPBD – perhaps the most important piece of legislation in Europe with respect to BEP assessments – choses to use average CFs as well [72,73]. Furthermore, Ref [73] also argues that most LCAs use average CFs. This is confirmed in our literature sample, in which only a single Ref that applies a LCP makes use of a marginal CF [33].

Another – albeit somewhat unusual – argument regarding average and marginal CFs is found in Ref [64]. In the context of charging EVs, the authors argue that a marginal CF should be used whenever *flexible* charging is taking place (in terms of the timing and the amount of energy charged), while an average CF should be used when the circumstances lead to the charging being completely *inflexible*.

Yet another argument revolves around which kind of CF is *easiest to calculate and use*. Refs [29,33] argue that a marginal CF has the advantage of being easy to use, because a (realistic) assumption can simply be made about which type of powerplant is most likely to be marginal. However, this ignores the complex reality of today's heavily interconnected electricity grid. Identifying the marginal generator in complex international networks is a highly non-trivial task. From this perspective, average CFs may in fact be easier to calculate and use, as argued by Refs [72,73].

None of the abovementioned arguments for or against a certain market perspective are particularly substantive or compelling. In fact, we can conclude from our review that the choice between average and marginal CFs is mostly a matter of perspective. Depending on the application, one or the other can be the preferred choice.

In any kind of *accounting method* that is meant to be applicable to all users (e.g. *all* buildings or *all* EVs), average CFs are the obvious best choice, because it is not feasible to consider all users as marginal users. Whenever the purpose is to calculate the *difference* made by any kind of demand-side intervention (e.g. manipulating the electricity consumption profile of a HP or an EV), marginal CFs may be preferred. Yet, even then some authors still prefer to use average CFs [36–38,41,47,58,74]. Potentially because the demand-side intervention is evaluated within a framework which itself uses average CFs. For example, when the demand response capabilities of a HP are considered within a traditional BEP assessment.

If one wishes to assess the 'difference made' by a very large-scale demand-side intervention (i.e. at the building stock level), an additional methodological choice with respect to CFs needs to be considered. Not only should a choice be made between average and marginal CFs, but also between exogenous and endogenous CFs. As explained further in section 3.4, it is important not to confuse these two choices.

In addition to marginal CFs, there are a number of other alternatives for average CFs. These revolve around CIs and their practical applications. A company may choose to apply a 'contractual CI' whenever it buys the rights to claim a particular (low) CI in the form of certificates that are generated by renewable producers (e.g. in the form of 'guarantees of origin'). Similarly, an individual home or EV owner may attribute a CI of 0 g/kWh$_E$ to his electricity consumption if he has signed a contract for 'green electricity' from a particular utility company. Contractual CIs are problematic for a number of reasons, including the fact that they do not credibly lead to *additional* renewable energy capacity and production, as discussed elaborately in Ref [83].

## 2.8. Import perspective

49 out of our 65 Refs (75%) use CFs that only consider the electricity production within the country (or countries) they focus on. In line with [76], we call this a *production perspective*. It ignores the fact that a large share of the electricity consumed within a country may be imported from abroad. Given the fact that the electricity generation mix in other countries is never identical to the local mix, these imports are bound to have an effect on CFs.

In 2017, eleven European countries imported more than 10% of their annual electricity consumption [84]. For several countries, the share of imports even exceeded 20%. Moreover, the share of imported electricity can even be much higher on an hourly basis. For example, Belgium has recently imported more than 50% of its electricity from abroad during certain hours [85].

The alternative to a production perspective is a *consumption perspective*. It does not simply ask "What are the CFs of electricity *produced* in this country?", but asks instead "What are the CFs of electricity *consumed* in this country?" The continually increasing interconnectedness between countries in the European electricity system potentially makes it increasingly important to apply a consumption perspective, even though it complicates the calculation of CFs.

Not taking into account cross-border flows of electricity has been recognized to lead to "significant distortions" [12]. It is therefore no surprise that a large number of Refs mention the desirability of taking them into account [9,12,21,26,32,37,42,45,72–77]. However, some of them only do so to highlight the fact that their production perspective is an important limitation [26,32]. They also note that applying a consumption perspective is a highly non-trivial thing to do.

We find that authors attempt to take imports into account in a variety of ways. A number of Refs take some kind of rudimentary approach. Ref [63] simply assumes a CI of 100 $g/kWh_E$ for all imported electricity, mentioning only that most electricity imported by Italy originates from France, where electricity is mostly produced with nuclear power plants. Refs [36,37] use the CIs published by the Danish TSO Energinet. Those take into account imports by assuming fixed CIs for imports from each of the countries neighboring Denmark (e.g. imports from Germany are *always* attributed a CI of 415 $g/kWh_E$, ignoring hourly fluctuations). Ref [45] calculates the CIs for electricity consumed in two US market zones. They use historical *price* data to estimate the marginal electricity generation technology and its associated $CI_T$. Since prices are co-determined by imports from neighboring market zones, imports are implicitly taken into account. It is important to note that such a price-based estimation would not work when *average* CFs need to be estimated.

Another group of Refs takes imports into account in a somewhat more sophisticated way. They consider historical data for both the electricity production in neighboring countries as well as how much was imported from those countries [9,42,71,74,77]. In Refs [71,77], the CFs for the country that is focused on have an hourly temporal resolution, but *yearly* average CIs are assumed for imported electricity from neighboring countries (similar to Energinet). Ref [74] also calculates hourly CIs for each of the neighboring countries, while emphasizing the added value of doing so. Ref [42] goes another step further by not only considering the electricity production in neighboring countries, but also considering the imports of those countries themselves (from other neighbors). To determine a CI for *those* imports, the authors simply assume the yearly average European CF.

Only a limited amount of Refs apply the *most* advanced methods to take into account imports. These methods consider the complex flows across the entire European network as a whole, instead of imposing an arbitrary geographical boundary (of one to two countries) around a specific country for which CFs are calculated. Refs [72,73] use a technique based on traditional MRIO models that are widely used in economics, referring to the earlier work of Ref [86]. This method is based on the balance between electricity production plus imports and electricity consumption plus exports, which is respected in every country node and timestep. Similarly, Ref [76] uses a method called 'flow-tracing', which the authors apply on historical data. For a particular historical hour, the CIs of each country are considered as vectors in a linear equation system, which can subsequently be solved. This process generates hourly CIs for each country, which take into account all flows across the network.

# 3. Discussion

## 3.1. Available sources of conversion factors

The fact that most of the Refs in our literature sample calculate CFs themselves can be explained by the fact that the externally available CFs are subject to a number of limitations. Table 3 provides a brief overview of these available sources.

*Table 3: Available sources of CFs*

| Source | CF | GS | TR | IP | AB | Refs | Limitations |
|---|---|---|---|---|---|---|---|
| IPCC | CI | EU 28 | y | p | o,l | [52] | Outdated (due to the slow publication cycle)<br>Highest temporal resolution is yearly<br>Does not consider imports |
| GEMIS | PEF CI | EU 28 | y | p | o | [87] | Outdated (based on EUROSTAT data 2010-2013)<br>Highest temporal resolution is yearly<br>Does not consider imports |
| EU JRC | CI | EU 28 | y | p | o,l | [88] [89] | Outdated (latest version published in 2017, refers to historical data from 2001-2013)<br>Highest temporal resolution is yearly<br>Does not consider imports |
| IEA | CI | EU 28 | y | p | o | [90] | Not publicly available<br>Simplified aggregations of certain technologies<br>Highest temporal resolution is yearly<br>Does not consider imports |
| EnergiNet | CI | DK | h | c | o | [91] | Only for Denmark<br>Simplified assumptions for CI of imports |
| RTE | CI | FR | h | p | o | [92] | Only for France<br>Does not consider imports |
| Tomorrow | CI | EU 28 | h | c | l | [93] | Historical dataset (since 2016) not publicly available |
| Ecoinvent | na | na | na | na | na | [94] | Only provides $PEF_T$ and $CI_T$ values for electricity generation technologies, not CFs for particular geographical areas |
| ENTSO-E | na | EU 28 | h | na | na | [95] | Only provides historical electricity production data for each electricity generation technology (on the basis of which CFs for particular countries can be calculated).<br>Some technologies with different $PEF_T$ and $CI_T$ values are aggregated (e.g. 'gas'), limiting the accuracy of CFs calculated on the basis of this data source. |

Note: **CF** = conversion factor used, **GS** = geographical scope of the CF (code of geographical areas like countries or the entire European area), **TR** = temporal resolution (yearly or hourly), **IP** = import perspective (production or consumption perspective on the CFs), **AB** = assessment boundary (is the CF calculated from an operational or a life-cycle perspective). Temporal scope and market perspective are not included in this table because all sources are retrospective and contain only average values. **Na** = not applicable.

The overview in Table 3 shows that – generally speaking – up to date, rigorously and transparently calculated CFs are not sufficiently available. This is problematic, because CFs are essential inputs for many kinds of academic and non-academic calculations, research and other applications. Our review also finds that this is a widely recognized problem. The insufficient public availability of CFs is frequently mentioned in our literature sample [37,39,73,76,77,83].

Notably, several recent studies have specifically focused on calculating CFs, as shown earlier in Table 1. But they do not meaningfully expand upon the sources shown in Table 3, because they have not made their full results publicly available. The exception to this is Ref [96], but its dataset of calculated CFs is limited to values for Switzerland, and is based on the outdated reference year 2015.

## 3.2. Transparency in conversion factor calculations

We find that full transparency about the calculation procedure behind CFs is unfortunately rare. In many instances, transparency is found to be weak or even completely lacking. Ref [97] calculates the PE and CE for a region of several municipalities in Italy, using the simulation tool EnergyPlan[3]. The authors mention that a PEF is used for electricity sourced from outside of the considered region (i.e. from the Italian grid), but they do not provide any explanation about which value is assumed or where it is sourced from. Similarly, the study reports CE results, but does not provide any information about which CI is assumed for electricity from the outside grid. Ref [25] calculates the PE associated with a ventilation system in a single-family building in Luxembourg, but only mentions briefly that the PEF is assumed to have a value of 2.7, without any source or further explanation. Ref [32] calculates the life-cycle PE and CE for a single-family building in Sweden. Here as well, the authors only mention the assumed CF values, without providing a source or any further information about the methodological choices behind them.

Weak transparency is common in the non-academic literature as well. National and local governments tend to use ad-hoc methodologies to calculate CFs and to measure progress on policy goals in terms of achieved reductions in PE and CE [6,59,98]. As noted by Ref [99], the CIs used by municipalities in the context of the *Covenant of Mayors* programme are often *self-declared*. Others explicitly use CIs from one of the sources mentioned in Table 3, which – as shown in the table – are often severely outdated and subject to other limitations like ignoring imports [99,100].

In the academic literature we reviewed, another aspect that is rarely reported on in a transparent manner, is which $PEF_T$ values were assumed. For most electricity generation technologies, the thermal conversion efficiency can simply be used to determine their $PEF_T$ value. However, some Refs seem to assume their own ad-hoc $PEF_T$ values. Ref [57] for example, assumes a $PEF_T$ of 1 $kWh_P/kWh_E$ for biomass, which – according to Ref [12] – is not even a recognized methodological option. For technologies like hydro, wind and solar, a methodological choice needs to be made to attribute either a value of 0 or 1 $kWh_P/kWh_E$ [12]. Presumably, most Refs that do not explicitly mention this choice, implicitly assume the most conventional value of 1. Others, like Refs [21,69] explicitly calculate PEFs using both the options of 0 and 1. Using those different PEFs to calculate the PE associated with the electricity consumption in their models, they confirm the importance of explicit $PEF_T$ assumptions. Many Refs only mention that their $PEF_T$ (and $CI_T$) assumptions were taken from the IPCC or EcoInvent databases, but this does not always clarify entirely the exact values used for each technology [26,40,42,74].

---

[3] www.energyplan.eu

At the time of writing, the CEN is developing a standardized protocol to calculate CFs [101]. Such a protocol is likely to leave open many methodological 'options' to its users, but at a minimum it will support academics as well as local and national governments to calculate their CFs in a more transparent way.

## 3.3. Limitations of using historical data to calculate conversion factors

Most Refs in our literature sample make use of historical electricity production data to calculate CFs. We find that the data used in many of these cases, is rather old. Eight Refs base their CFs on historical data referring to electricity production that took place four years before they were published [2,27,30,62,64,66,73,74]. Twelve Refs make use of data that is five or more years older than the publications themselves [9,10,31,34,50,52,53,55,56,58,70], which is 21% of all Refs that make use of CFs based on historical data. Notably, three of those Refs make use of data that is even ten or more years older [10,52,53]. In other words, many authors risk their PE and CE results to be immediately outdated upon publication.

In addition to the frequent problem of outdatedness, using historical data to calculate CFs is subject to the limitations inherent to those data. For example, historical datasets by the IEA and ENTSO-E aggregate several electricity generation technologies into categories like 'gas', while the technologies contained within these categories each have different $PEF_T$ and $CI_T$ values. In some cases, these aggregations are even as broad as to group all '*fossil'* electricity generation, ignoring extensive internal differences [67]. This can naturally lead to imprecise PE and CE calculations.

## 3.4. Endogenous conversion factors

All Refs presented in Table 1 use exogenous CFs. This means that the CFs they use are assumed to remain unaffected by the consumption of electricity that is being considered. For most use-cases in Table 1, this is an obvious methodological choice. When calculating the PE or CE related to an individual appliance, building or EV, it is reasonable to assume that the examined electrical loads do not meaningfully affect the large-scale electricity system. Put differently, the 'ideal use-case' for exogenous CFs is precisely these kinds of calculations. For the Refs in question, any methodological concerns relate primarily to the different aspects discussed in section 2. Not to the fact that exogenous CFs are used.

For other kinds of calculations, the use of exogenous CFs is more debatable. Namely, when PE and CE calculations are performed for electrical loads on a much larger scale. For example, when instead of considering an individual HP, renovation measure or EV, their large-scale *roll-out* at the societal level is considered. These kinds of *'roll-outs'* (as we henceforth refer to them) are also studied widely in academic literature [78,102–108].

If roll-outs cause a significant change to the societal electricity demand, the electricity system (i.e. the supply side) will have to adapt. Exogenous CFs may then no longer suffice. Instead, it may be desirable to model the electricity system as well, to estimate how CFs themselves are affected. Studies that do this, effectively use 'endogenous CFs'. A term we use for the sake of a clear taxonomy. In practice, these studies usually calculate the PE and CE associated with a roll-out by observing how the PE and CE in their endogenously modelled electricity system changes. Therefore, they do not explicitly use or mention any CFs, which is why they are excluded from Table 1 and from our 'structural review' in section 2.

A major limitation of using endogenous CFs, is the fact that it typically requires a number of simplifications to be made. This results from the fact that it is highly non-trivial to rigorously model both the demand and supply sides. Let alone in a completely integrated way, allowing for complex interactions. All while keeping the integrated model computationally manageable.

On the supply-side, this usually results in a simplified representation of the electricity system, compared to state-of-the-art electricity system models [102–104]. Common simplifications are a (sharp) reduction in the number of electricity generation technologies, and ignoring imports from interconnected electricity systems. Meanwhile, the demand-side (where the roll-out itself takes place) may be heavily simplified as well. For example, by determining the hourly space heating demands of an entire building stock solely on the basis of heating degree data, ignoring other variables like behavior [107]. With respect to calculating the PE and CE associated with roll-outs, these kinds of simplifications potentially counteract (some of) the gains in accuracy that are hoped for when using endogenous CFs.

It should also be noted that even endogenous CFs are subject to the methodological boundaries of the underlying models themselves. For example, consider a study using endogenous PEFs, which calculates the PE reductions associated with a large-scale renovation of buildings. Such a study is likely to ignore the fact that the PEF may simultaneously be affected by the electrification taking place in the transport and industrial sectors, over the same time period in which the renovations take place.

Even if all sectorial electrification is considered in unison, the optimization approaches that are typically applied in these studies remain subject to certain methodological boundaries. For example, the fact that – in reality – the electricity sector is largely influenced by political decisions, instead of resembling a purely techno-economic optimization.

Some studies therefore choose to apply exogenous CFs, even when large-scale roll-outs are being considered. Four such studies can be found in our literature sample [57–60]. They each calculate an amount of PE or CE at the level of an entire building stock. The use of exogenous CFs then allows for other modelling aspects (on the demand side) to receive a greater amount of attention. Instead of endogenizing the impact on the electricity system, authors could – for example – invest primarily in a more realistic BEP model, to better assess heating requirements before and after renovation. The BEP model could then include a multi-zonal heating representation, behavioral aspects like rebound effects, or a more detailed representation of technical installations like HPs. Similarly, authors may want to pay attention to life-cycle aspects, which potentially require significant additional modelling efforts. Research attention paid to these other aspects can be 'crowding out' the endogenization of the electricity system. However, their added value may be of a similar or even greater magnitude. Choosing for exogenous CFs in the context of analyzing roll-outs can therefore be entirely legitimate.

Finally, it is important to stress that the choice between exogenous and endogenous CFs should not be confused with the choice between average and marginal CFs. For example, when certain authors claim that average CFs should be used when demand response by buildings is analysed on a small-scale, while marginal CFs should be used when the demand response takes place on a large scale [75]. A distinction between analyses on a small or large scale more appropriately calls for a consideration of exogenous versus endogenous CFs.

## 3.5. PEFs in the context of official BEP assessments

The concerns raised in our literature review apply outside of the academic realm as well. Across Europe, PEFs are used in the calculation methodologies for official BEP assessments. Its value co-determines how buildings – as well as the associated technologies and policies – are evaluated. Therefore, it has been the subject of considerable debate.

The 2012 Energy Efficiency Directive (EED) proposed a European PEF of 2.5, based on an assumed 40% average thermal efficiency in the European electricity system [109]. In 2016, the European Commission procured a study to evaluate the PEF, which concluded that the value should be updated (lowered) to better reflect the current state of the European system [12]. The 2018 revision of the EED confirmed this by stating that the PEF "should be reviewed", although a new value was not proposed in the Directive itself. Therefore, the outdated value of 2.5 is still being widely used.

Ideally, Member States would adopt national PEF values that are up-to-date and better reflect national circumstances than a European average. They are allowed to do so, but they are unfortunately not obliged to be transparent or to use a particular methodology [109]. Therefore, it is unclear how the various national values that are already in use across Europe were calculated, and whether or not they can be mutually compared in a consistent manner [6]. Potentially, the adoption of the CEN prenorm prEN17423 (in its finalized form) could alleviate this problem.

However, the finalized norm will leave many 'options' open to the Member States. This could reinvigorate a debate about the methodological aspects discussed in section 2. While the European PEF value of 2.5 is purely retrospective in nature, applies a yearly temporal resolution, and applies an operational perspective, each of these options could be reconsidered in the process of adopting up-to-date national PEFs. After all, using prospective values, a higher temporal resolution and a life-cycle perspective can each have a considerable added value.

In the context of official BEP assessments, prospective PEFs could enable the PE associated with a particular building to be projected for a period of 10 or 20 years. This would better approximate the PE that should actually be expected from a building in its foreseeable future, compared to the current (implicit) assumption that the PEF will remain constant over time. It is possible that a building that does not reach a regulatory PE target under a constant (retrospective) PEF, *does* reach the same target when a prospective PEF is used. Meanwhile, hourly PEFs could better take into account the PE associated with temporal fluctuations in a building's electricity demand. This may be especially useful when a building's hourly electricity demand fluctuates heavily, for example when a HP is combined solar PV.

Adopting prospective and hourly PEFs in official BEP assessments is entirely feasible in principle. Values for any Member State can be calculated using the available tools to model the European electricity system (e.g. PRIMES and METIS)[110,111]. A standardized set of calculated values could be used across BEP assessments, in line with the well-established practice of using standardized values for outside temperatures and solar irradiation [6,7][4].

It is not yet clear how the adoption of prospective and hourly PEFs would affect official BEP assessments in practice. To properly study these impacts across Europe, an up-to-date database of prospective and hourly PEFs (covering the various Member States) would be required. However, such a database is not available at the time of writing.

[4] One caveat would be to make sure that all weather data used in a BEP assessment is synchronized to the same historic year, including the calculation of the CFs themselves.

## 3.6. Conversion factors in the context of developing smart grid algorithms

Another application of CFs, especially when hourly values are available, is the development of smart grid algorithms (SGAs). SGAs are techniques to optimally schedule electrical loads from individual appliances, entire buildings or EVs[5]. Traditional challenges in the development of SGAs include dealing with uncertainty and combining optimization objectives. In our literature sample, we find many Refs that take part in SGA development in one way or the other [24,27,28,35–38,47,56,73,112]. Each of these Refs explicitly uses either PEFs or CIs as (one of the) control signal(s) in their SGAs. By helping to reduce the PE and CE associated with various electrical loads, SGAs can make an important contribution to the realization of the European climate and energy policy goals.

Notably, Refs [36–38] make use of the hourly CIs published by the Danish TSO EnergiNet. However, such hourly CIs are generally unavailable in other countries. As noted by several authors, larger and better datasets of hourly CFs could provide a crucial impetus for the further development of SGAs [67,72,73].

*Prospective* hourly CFs would be especially useful in this context, because the projected increase in VRES is expected to increase the volatility of hourly CFs. As noted by Ref [27], this volatility is a crucial parameter determining the potential reductions in PE and CE that can be achieved with SGAs. Projecting hourly CFs into the long-term future would therefore enable an assessment of how the performance of SGAs could be affected as the penetration of VRES continues to increase. In other words, the unavailability of a dataset containing hourly prospective CFs for (most or all) European countries is an impediment for the development of SGAs as well.

In practice, the implementation of SGAs will eventually require prospective CFs that consider the short-term future. Continuous forecasts of CFs could feed the SGAs for smart appliances, buildings and EVs across Europe. For example, to focus their electrical loads on periods with a low CI. Producing these forecasts lies entirely within the realm of possibility, since ENTSO-E already continuously produces 72h forecasts of the hourly electricity production across Europe, as noted by Refs [72,73].

[5] SGA is a generic term we use for the sake of clarity and conciseness. It is not necessarily used in the associated Refs themselves.

# 4. Conclusion

PEFs and CIs play an important role in the context of the European energy and climate policy goals. They help track progress towards the goals, and they enable comparisons between different technologies and measures that can play a role in reaching them. More specifically, they co-determine the PE and CE that should be associated with individual appliances like heat pumps, entire buildings (before and after renovation measures) and EVs. As the heating and transport sectors continue to electrify, the importance of CFs will only increase in the coming decades.

To the best of our knowledge, we performed the first structural review of how CFs are actually calculated and used in recent academic literature. For each of the 65 Refs included in our literature sample, we evaluated six methodological aspects concerning CFs.

For four of the methodological aspects, we found that a majority of Refs chooses for what is arguably an inferior approach (Figure 1). Namely, to only consider the CF of a single country (72%)[6], to apply a purely retrospective perspective (86%), in which case, historical data that is five or more years old is often used (21%), to apply a yearly temporal resolution (54%), and to ignore electricity imports from other countries (75%). This stands in contrast to the fact that many Refs across our literature sample have shown that considering the geotemporal variability of CFs (including in the foreseeable future) and considering imports can each have an important added value when calculating the PE or CE associated with a particular electrical energy demand.

The arguably superior options for all of the abovementioned aspects can in principle be combined without causing additional methodological hurdles, with the exception of taking imports into account and calculating CFs for several countries. To accurately calculate CFs from a consumption perspective for a large number of countries, advanced methodologies are required. Only three of the Refs in our literature sample used such methodologies. None of them have calculated prospective CFs.

Prospective CFs could provide a particularly significant added value, because they would allow PE and CE to be forecast for a number of years. For example, calculating the expected CE that should be associated with a particular heat pump or electric vehicle across the coming ten years. Moreover, if prospective CFs also have a high temporal resolution, expected trends in their seasonality and hourly variability could be quantified, as well as the impact of these trends on PE and CE calculations.

In our review of the literature, we also found that most Refs use average CFs (91%) instead of marginal CFs, and that most Refs apply an operational rather than a life-cycle perspective (65%). In the case of average versus marginal CFs, choosing a superior option is highly debatable. One can defend the use of marginal CFs in the context of certain applications, but they do come with a number drawbacks and complications. Marginal CFs have this in common with endogenously calculated CFs, which they are not to be confused with. In the case of applying an operational versus a life-cycle perspective on CFs, the latter can ultimately be seen as the superior approach. However, the associated additional effort and assumptions may seem excessive for certain applications. Regardless of the approaches taken in future research, authors should always be aware of all six of the abovementioned methodological aspects.

[6] In narrow cases where the research interest is purely focused on a single country, applying only that country's CF may not be considered an inferior approach. However, applying the CFs of other countries is a superior option in the case of most academic studies, because it reflects on the sensitivity of results to the applied CFs and makes the results more generalizable. This may however require some additional effort, as discussed in section 2.3.

A few of the Refs in our literature sample have focused exclusively on the calculation of CFs itself. However, none of them have made their calculated CFs publicly available (with the exception of a single Ref which calculated CFs for a single country in the year 2015). A number of sources for PEFs and CIs are publicly available, but each of them is either severely limited geographically (e.g. the Danish TSO's publication of hourly CIs) or heavily outdated (e.g. the IPCC, GEMIS and EU JRC datasets). Moreover, none of them include prospective CFs.

An ideal dataset would contain CFs that have each of the desirable characteristics described above, calculated in a coherent and transparent manner. This would be useful for a number of reasons. First of all, the ideal use-case would be the calculation of the PE and CE associated with individual electrical energy demands. Secondly, exogenously calculated CFs can legitimately be used to calculate the PE and CE associated with large-scale electrical energy demands (e.g. for entire building stocks). Third, it would allow for a critical evaluation of the regulatory CFs used in official BEP assessments. Fourth, it would allow for a meaningful *comparison* of CFs across European countries, in terms of yearly averages, seasonality effects, hourly volatilities and the degree to which CFs are co-determined by imports. Both now and in the foreseeable future. Finally, it would also support the process of developing *new* BEP assessment methods and SGAs, which can help contribute to the European energy and climate policy goals. Nonetheless, the described dataset remains unavailable at the time of writing.

As noted by the landmark study procured by the European Commission in 2016, generating this type of dataset would require "very complex power sector model calculations which […] would have to be carried out using a highly detailed European model." [12, p. 39]. Given the strong need in the literature for the described dataset, we develop such a European model in our future work. The model will be used to generate the desired dataset, which we will analyze and make publicly available for all academic and non-academic purposes described above.

# 5. Acknowledgement

We gratefully acknowledge the financial support received for this work from the Fund for Scientific Research (FWO) in the frame of the strategic basic research (SBO) project "NEPBC: Next generation building energy assessment methods towards a carbon neutral building stock" (S009617N).

# 6. Appendices

## Appendix A: reasoning behind the categorization of Refs in Table 1

In the case of the building-related categories in Table 1, the following reasoning was applied to distinguish them. Ref [46] was put in the '*multiple buildings'* category, because its PE and CE calculations, strictly speaking, only cover a few individual buildings, even though these buildings are claimed to be representative of an entire building stock. Following the same line of reasoning, Refs [59,60] were put in the '*building stock*' category, because they *actually* make PE and CE calculations *for entire building stocks*, even though they do so by extrapolating from a small number of representative buildings. Similarly, Ref [2] is put in the '*multiple buildings*' category instead of the '*municipalities'* category, even though the selection of buildings for which the PE and CE are calculated are claimed to be representative for particular municipalities.

## Appendix B: Supplementary information accompanying the literature sample

*Table B.1: Supplementary information about the Refs included in our literature sample*

| Ref | Note |
|---|---|
| **Single appliance level** | |
| [19] | Yearly average CI calculated by linearly extrapolating from the latest available historical value of each of the four countries, down to zero in the year 2050. Finds that the battery can increase solar PV self-consumption and reduce both the CE and peak demand associated with the HP. |
| [20] | 'Current' PEF calculated on the basis of historical data (2014). 'Future' PEF calculated by adjusting wind and solar capacities to the values found in external scenarios (from ENTSO-E) and performing a simplistic estimation of the capacity factor of other technologies, given those renewable capacities. PE savings are estimated on the basis of changes in the consumption of gas in (a) buildings and (b) the electricity system. |
| [21] | Assumes both a $PEF_T$ of 0 and 1 for VRES (calculating PE in each case). |
| [23] | Electricity system representation focusses on the responsiveness of fossil generators to changes in electricity demand, as observed in historical data. |
| [24] | Also performs an integrated analysis at the national level (i.e. taking into account the impact on PEF if there would be many such HP + TES installations). The same endogenously calculated PEF is used in the analysis at the individual appliance level. |
| [11] | CI value is based on historical data (EEA calculation for the year 2014). Several PEFs are used; the traditional European average value of 2.5 established in the 2012 EED, the 2.0 'updated' value proposed by [12] and a value of 1.8 which is assumed to be a good estimate of the future PEF. |
| [26] | Purpose of the study is to identify the difference in outcome when hourly CIs are used instead of yearly CIs. |
| [27] | Operation of the HP is optimized to minimize CE on an annual basis, by using the hourly CI as a control signal. The CI calculation method is analogous to Refs [72,73]. |
| **Individual building level** | |
| [28] | Estimates the possible reduction of the CE associated with a residential building's total electricity consumption, through the optimization of the operation of the home battery. |
| [29] | Potential renovation scenarios for a particular building are simulated. For each scenario, PE and CE related to the building's electricity demand is calculated. PEF and CI are 'calculated' by assuming that a coal plant is always marginal in Sweden, and that coal plants have a thermal efficiency of 35%. |
| [1] | PE and CE calculated for a variety of renovation scenarios for the building in question. PEF values based on a governmental report, the values of which are based on historical data. The building is thought to be representative of many such typical apartment buildings across Finland and comparable geographies. |
| [30] | Building design is optimized on the basis of both costs and achieved PE. The building contains a HP for both heating and cooling. PEF provided in Italian regulation is used, which is based on historical data. |
| [31] | Nearly identical analysis as Ref [30]. Building design is optimized on the basis of both costs and achieved PE. The building contains a HP for both heating and cooling. PEF provided in Italian regulation is used, which is based on historical data (one year older data than Ref [30]). |
| [10] | Many renovation scenarios are simulated, including scenarios with HPs. PEF and CI are taken from a 'Czech GEMIS database' which is based on historical data. As a sensitivity, PEF and CI values are increased and decreased by 10%, 20% and 30%. In the scenarios containing a HP, this generates '*substantially different results*'. |
| [32] | The building only consumes electricity. It is heated with a HP. Full LCA, considering not only the electricity consumption of the building, but also its embodied energy and emissions. |

[33] The building is connected to a DHN, so the electricity consumption for which the PE and CE are calculated is non-heating related (e.g. ventilation). PEF and CI is set to the $PEF_T$ and $CI_T$ of a coal power plant, which is assumed to always be the marginal plant in the Swedish system. As a sensitivity, the values are changed to the $PEF_T$ and $CI_T$ of a gas power plant, which is assumed to always be the marginal plant in the future Swedish system.

[34] Analysis of a retrofit scenario which reduces the buildings' PE to the NZEB level. Italian national PEF is taken from an Italian Ministerial Decree, which is based on historical data.

[35] The authors use PEF and CI values based on historical data, calculated by Ref [71]. The building's battery system uses the hourly CI values to minimize the buildings' CI.

[36] CE of the building is minimized though MPC that uses the hourly CI signal published by the Danish TSO EnergiNet.

[37] Cf. [36], with the difference that the historical data for the year 2015 is used.

[38] Cf. [36], with the difference that the historical data for the years 2013 and 2014 are used, and an apartment building is considered (the electrical heating of which is optimized to minimize CE).

[39] The buildings' HP is used both for heating and cooling. MPC is used to minimize CE.

[40] Explicit purpose is to find a difference in outcome when yearly and hourly CIs are used.

[41] Hourly simulation explores opportunities for EVs to reduce the buildings' CE and PE, but yearly average CI and PEF are used. Building CE and PE are reduced by increasing the self-consumption of its locally produced solar energy. EV CE and PE are reduced by being (partially) charged with the buildings' solar energy, as compared to (a) a case with an EV being charged purely from the grid and (b) a case with an ICE, with its respective CE and PE.

[42] PEFs are separately calculated considering only the renewable electricity generation, or only the non-renewable electricity generation. The building only consumes electricity and is equipped with a HP and a PV installation.

**Multiple buildings level**

[43] Building electricity demand for which PE and CE is calculated, is limited to 'non heating related' demand. Current and future PEF and CI value sourced from an external governmental report, which considers historical data and makes a linear extrapolation into the future (2015-2027).

[45] CI is estimated on the basis of historical electricity *price* data, which is assumed to represent various marginal electricity generation technologies at each price level.

[46] Buildings are representative of the Portuguese building stock. Authors use the PEF value determined by the Portuguese building energy performance regulation, which is itself based on historical electricity generation data.

[2] Building samples are representative for the analyzed Swedish cities. Three CI values are calculated and applied, each representing the 'the average CI in Nordic countries' in a different historical year.

[47] Hourly CIs and PEFs calculated on the basis of historical data from the Belgian TSO Elia. One building is heated with traditional electrical resistance heating, the other with a heat pump. Both make use of the TES capacity of their floor heating systems. In separate simulations, the CE and PE of both buildings is minimized through MPC that uses the hourly CIs and PEFs signals. Stylized prospective PEFs and CIs are calculated by extrapolating VRES production to 40% of demand.

**Municipality level**

[48] Calculation uses Spanish national CI based on historical data, found in a report from the Spanish Association of the Electrical Industry.

[49] Calculation uses Croatian national CI based on historical data, found in a report from the Croatian Ministry Of Environment And Energy.

[50] Calculation uses Italian national CI based on historical data, found in a report from the Italian Ministry of Environment.

[51] Calculation uses Italian national CI based on historical data, found in a report from the 'National Emission Inventories'.

[52] Performs the CE calculation both with the Italian national CI and a more local CI for Sicily, to compare the outcomes. Calculations are performed both on from an operational and an life-cycle perspective. The operational national value if sourced from the IPCC (based on historical data, 2006). The LCA national value is sourced from the ELCD (based on historical data, 2008).

[53] Calculation uses Spanish national CI based on historical data, found in a report from the IPCC.

[54] Calculation uses Croatian national CI based on historical data found in a report from the Croatian Ministry Of Environment And Energy.

[55] Calculation uses Greek national CI based on historical data found in the 'CoM Guidebook' made by the EU JRC [88].

[56] Hourly CIs are *partially* based on historical electricity production data for renewables and nuclear. The electricity generation by technologies for which data was not available, is estimated with a capacity and dispatch optimization model to cover the residual demand.

**Building stock level**

[57] Danish 2050 electricity system is simulated in the energy-system tool EnergyPlan. PEFs are calculated based on the results from EnergyPlan. Biomass is assumed to have a $PEF_T$ of 1, while all other electricity generation technologies in the 2050 system (i.e. other renewables) have an assumed $PEF_T$ of 0, resulting in a PEF of 0.2.

[58] External scenarios used to project the future evolution of the various building stocks. Scenarios include varying levels of heat supply electrification. National CI values are calculated on the basis of historical data and *are assumed to remain constant* in the future scenarios considering the year 2030.

[59] The impact of various renovation measures on building stock CE is calculated. Renovations take place on $t_0$ and CE is calculated for a number of years into the future. CIs are sourced from government reports, which are based on historical data. The historical year on which the CIs are based, varies from country to country (depending on the report used). CIs are assumed to remain constant across the future years considered (2030 and 2050).

[60] CIs and PEFs are sourced from an external report, the values of which are based on historical data. The synthetic building stock is meant to represent Switzerland in the year 2015.

**Electric vehicles**

[62] CI is calculated on the basis of historical data from the Belgian TSO (Elia).

[9] Yearly average CI per country based on IEA historical data. This is combined with ENTSO-E historical data to take into account imports.

[64] CI calculation is based on historical data from grid operators in France and Germany.

[65] CIs for both Poland and Czechia are taken from IEA reports that estimate the electricity production mix on a yearly basis from 2015 up to 2050 (in 5 year steps). On top of this, one LCA analysis is also performed with the assumption that EV's can be fully charged by renewables only, enabled by a smart grid environment (cf. "o" in the GS column).

**Other references**

[67] The used hourly historical data problematically aggregates various thermal electricity generation technologies. To deal with this, available *yearly average* data are used to disaggregate the hourly data into a more detailed mix of technologies. This approach is pragmatic but not does not produce ideal data to accurately calculate PEFs or CIs. Moreover, their disaggregation still leaves 'gas' as a single generic technology. I.e. it makes no distinction between peak (OCGT) and non-peak (CCGT) gas power plants, which have a different thermal efficiency and thus a different $CI_T$ and $PEF_T$.

[68] Authors claim that nuclear $PEF_T$ should be as high as 60 $kWh_P/kWh_E$ if the remaining PE in nuclear waste is taken into account. Based on historical data, this would increase Swedish PEF from 1.8 to 25.5 $kWh_P/kWh_E$ and European PEF from 2.5 to 18 $kWh_P/kWh_E$.

[70] The word 'dynamic' in the Ref title does not refer to a high temporal resolution (e.g. hourly), but to the fact that yearly values can be "easily updated" on a year-by-year basis, through the proposed methodology.

[71] Renewable and non-renewable PEFs are separately calculated.

[72] Includes a brief discussion about the potential for smart buildings to make use of hourly CI signals.

[73] Includes a limited application of the hourly CIs on a Norwegian single-family building.

[74] Includes a rudimentary estimation of how much CE could be reduced when a theoretical load consumes 1kWh every day during the hour with the lowest CI.

[75] European yearly average CIs are calculated with an UCED model based on scenarios developed in the EU funded project SUSPLAN (for the years 2010, 2020, 2030, 2040 and 2050).